# Evidence for Energy Regularity in the Mendeleev Periodic Table

C. H. S. Amador[1)1] and L. S. Zambrano[2)]

[1)]Centro Brasileiro de Pesquisas Físicas
R. Dr. Xavier Sigaud, 150; Rio de Janeiro CEP 22290-180   Brazil
[2)]Universidad de Puerto Rico, Mayagüez, Puerto Rico

Abstract

We show that the dependence of the total energy of the atoms on their atomic number can follows a q-exponential (as proposed by C.Tsallis), for practically all elements of the periodic table. The result is qualitatively explained in terms of the way the atomic configurations are arranged to minimize energy.

One of the greatest scientific achievements in History was the elaboration of the periodic table by Mendeleev. Indeed, it is the most successful scheme to describe the organization of chemical elements and, after its physical structure was unveiled by Bohr in terms of electronic configurations, it became the basic working tool not only in chemistry but also in atomic and molecular physics [1]. The electronic structure of the atoms in their ground states, which determines their position in the periodic table, is a state of minimum energy in the Coulomb field, with all mutual interactions taken into account, subjected to the rules of Quantum Mechanics. However, as already noted by some scientists, the energy of the atoms is a parameter not present in the table. In the words of Allen, the periodic table "*is the most powerful instrument for organizing chemical phenomena but it does not contain any information about the energy of the atoms*" [2].

Actually, the energy of the atoms in their ground states, i.e., the sum of the energies from all the occupied electronic levels, is an essential parameter, in particular to gauge the accuracy of the variational methods used to calculate the electronic configurations [3-5]. It follows from the calculations carried out by many different authors, using the ab-initio Hartree-Fock (HF) and Density Functional Theory (DFT) methods, that the variation of the value of energy with atomic number follows a simple monotonic curve, shown in Fig.1. The data consolidated in this curve runs over all elements of the periodic table, from hydrogen to lawrencium, obtained trough a generalized Gaussian basis set with HF [6]. Since the energy of an atom, besides the simplest ones, depends on rather complex arrangements of the electronic levels, one cannot avoid questioning why the dependence with atomic number seems that simple and whether an analytical expression could be found to express this dependence without complex calculations.

---
[1]   The current author adress is Instituto Superior Técnico, Av. Rovisco Pais, 1049-001 Lisboa, Portugal

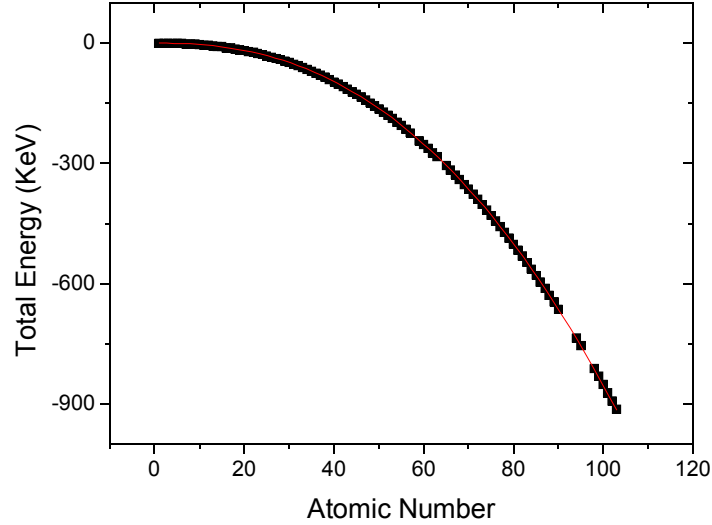

Fig. 1. Total energy of free atoms in ground state from ¹H to ¹⁰³Lr (fitted with eq. (3)). [6]

Starting with the second half of the question, obviously one can use an analytical expression, such a polynomial, to fit the data represented in the curve of Fig. 1. However, as there is not a unique curve to fit the data, and to correct for the non-normality of the residuals of simple fittings, we decided to use a statistical approach, the Box-Cox method. This procedure is designed such that we can compare the effect of various power transformations of the energy on its linear regression with atomic number [7]. Leaving aside technical details, we found out that the data can be best fitted by an expression that can be readily converted into a q-exponential [8,9]:

$$E = E_H [1 + B(1-q)(Z-1)]^{\frac{1}{1-q}} . \qquad (1)$$

Having this result, we can turn to the first part of the question.

The atoms can be considered as a system of charged particles interacting through electromagnetic forces and subject to the constraint imposed by the Pauli Exclusion Principle. The dynamics of the system is governed by the Schrödinger Equation, so that the equilibrium configuration depends on the energy levels given by its stationary solutions. Therefore, when considering energy minimization procedures, which form the basis of all variational methods, one realizes that the quantum rules imply that there exist strong correlations between subsets of the interacting particles, so that the minimization path has to satisfy them before seeking the overall minimum energy of the entire system.

Kodama and collaborators have shown that the probability distribution function of complex systems following this scenario, i.e., with strong correlations between particles of any single-particle state, follows the Tsallis q-exponential [10]. Clearly, these considerations suggest analyzing the data of Fig. 1 in terms of this function.

This can be mostly conveniently done by plotting the q-logarithm [8],

$$\ln_q Y = \frac{Y^{1-q} - 1}{1 - q}, \quad (\ln_1 Y = \ln Y) \tag{2}$$

of the energy ratio $Y = E/E_H$, where $E_H = -13.60534$ eV of the hydrogen atom, as a function of the atomic number $A$. The proper value of $q$ can then be consistently found by imposing that the resulting curve be a straight line with unity correlation. This is shown in Fig. 2 for different values of $q$.

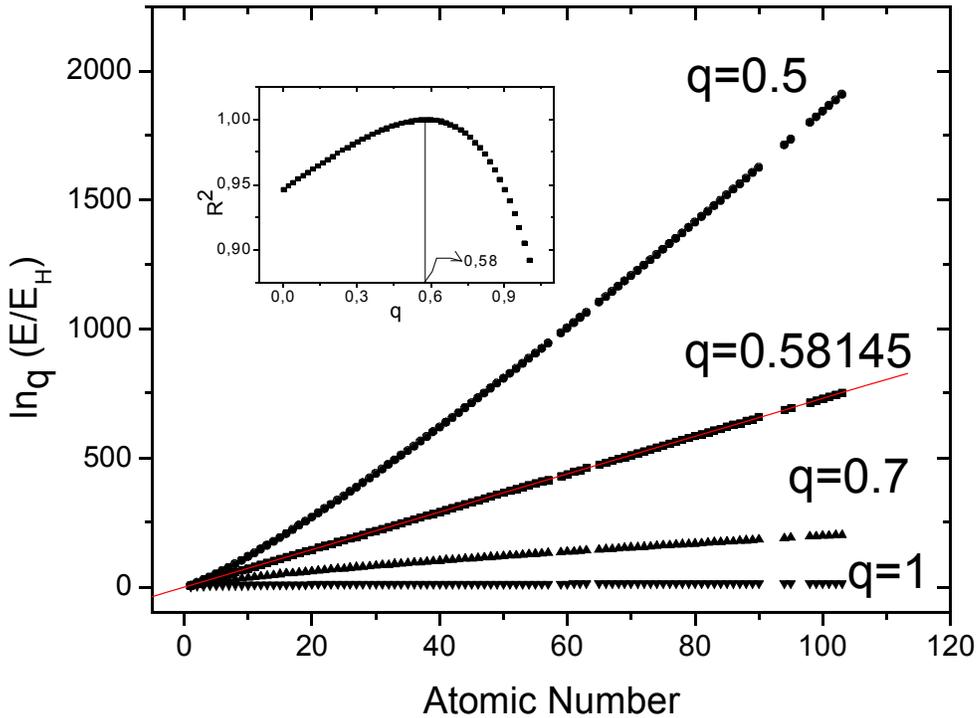

Fig. 2. The data of Fig. 1 plotted in terms of the q-logarithm of the ratio of the energy of the various atoms to the energy of hydrogen, as a function of the atomic number and for different values of $q$. The insert shows the correlation function of the fitting of different curves by a straight line. The optimum value of $q$ is given by $R^2 = 1$ (6-digit precision).

The optimum value of the $q$ parameter is found to be $q = 0.58145$, so that we obtain the quite interesting result that the energy of the ground state of all atoms of the periodic table is given by eq. (1), with $q = 0.58145$ and $B = 2.4333$, i.e.,

$$E = E_H\left[1+1.0185(Z-1)\right]^{2.3892} . \qquad (3)$$

This result is remarkable in that it holds for the entire periodic table with unity correlation. The above mentioned Box-Cox method leads to the same result.

The somewhat unexpected effectiveness of the q-exponential, to represent the energy of the elements of the periodic table, hints to the possibility that it is also capable to represent the energy of more complex atomic systems, in which the electronic configuration resembles that of isolated atoms. One such a system that is particularly relevant is the encapsulation of atoms into hollow fullerenes [11,12]. It has been experimentally demonstrated that nitrogen implantation, for instance, "*produces a paramagnetic center with hyperfine interaction properties very close to that of atomic hydrogen*"[11]. Furthermore, the endohedrally doping $C_{60}$ with different atoms, such as H, He, La, Cu, is being intensively investigated with respect to interesting physical properties, such as high-temperature superconductivity [12,13]. In these compounds, the doping atom is trapped inside a carbon nanocage by the potential wall produced by the fullerene electronic cloud. Considering the origin at the center of the fullerene, this cloud exists between 2 and 5 Å, with a maximum around 3.5 Å.

The total energy of doped fullerenes has been calculated for 18 different doping atoms (the covalent atoms $^6$C, $^7$N, $^8$O, $^9$F, $^{14}$Si, $^{15}$P, $^{16}$S, $^{17}$Cl, and $^{35}$Br, and the transition metals $^{21}$Sc, $^{22}$Ti, $^{23}$V, $^{24}$Cr, $^{25}$Mn, $^{26}$Fe, $^{27}$Co, $^{28}$Ni, and $^{29}$Cu), using the density functional method, with B3LYP exchange and correlation term [14] and 6-31G* basis set [4]. The total energy of the corresponding compounds as a function of the atomic number of the doping atom, placed at the center of the carbon nanocage, is shown in Fig. 3, which clearly resembles Fig. 1. Discounting the energy of the fullerene without doping, which is 62.21 keV, the results can again be quite well represented by the q-exponential given by Eq. (1), with the value of $E_H$ substituted by $E_{FUL}$ = -13.726 eV (with a $10^{-3}$ precision), with a $R^2$=1.00000.

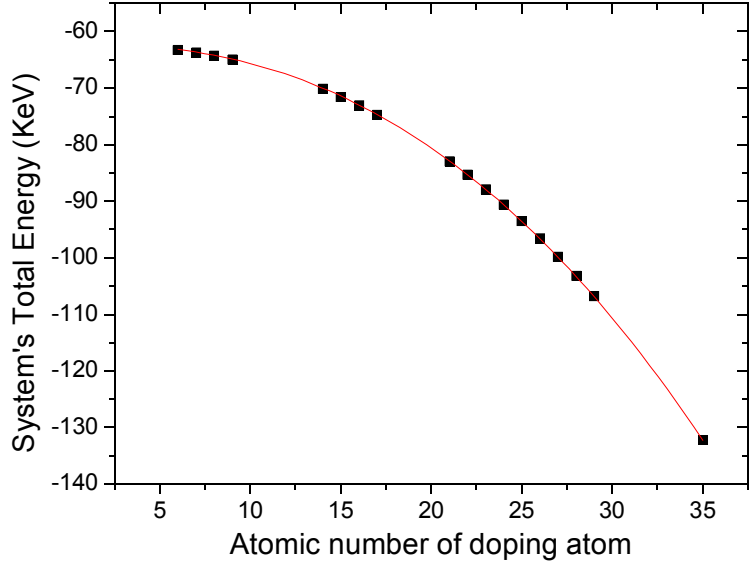

Fig. 3. Total energy of $C_{60}$ doped with different atoms trapped at its center as a function of the doping atom atomic number. $R^2 = 1$ (6-digit precision).

This interesting result can be understood as if the fullerene provided a background energy to the energy of the doping atom, as the function depends only on its atomic number and not on the atomic number of the complete system.

To compare these results with existing expressions for total energy in ab-initio methods is not simple. Nevertheless, we note that within the Thomas-Fermi approximation, which was the first one to use the electronic density as the main parameter, one could find, through heuristic arguments and for large Z (eq. (11.1) in ref. [15]):

$$E[Z] = -\left(c_7 Z^{7/3} + c_6 Z^2 + c_5 Z^{5/3}\right) Ry + ... , \quad (4)$$

where $Ry$ is a Rydberg ( $-E_H$ in our expressions), $c_7$=1.53749024, $c_6$=1, and $c_5$=0.5398. Using the same limit in eq. (3) and taking only the first term, we have:

$$E = -1.0448 \ Z^{2.3892} Ry . \quad (5)$$

It is clear that this result resembles the first term of eq. (4) (7/3 $\simeq$ 2.333) and indeed one does not expect exact agreement, because the Thomas-Fermi method is an approximation. Other expressions similar to eq. (4) can be tried; however they work only for rather large values of Z.

We realize that a derivation of the q-exponential for the energy from first principles is lacking and that the value q = 0.58145 may not be widely valid. Nevertheless, we believe that our result will certainly stimulate further research in this direction.

**Acknowledgments**

To Capes and CNPq for financial support, C. Tsallis for discussions and suggestions, C.A. Taft for computer simulations and R.M.O. Galvão for text revision and insights.